\documentclass[fleqn,10pt]{wlscirep}
\usepackage[utf8]{inputenc}
\usepackage[T1]{fontenc}
\usepackage[symbol]{footmisc} 
\usepackage{siunitx}
\usepackage{booktabs}
\usepackage{wrapfig}

\newcommand{\set}[1]{\left\{#1\right\}}
\newcommand{\dist}{\mathrm{d}}
\DeclareMathOperator{\arcosh}{arcosh}
\DeclareMathOperator{\atan2}{atan'}

\title{The hyperbolic geometry of financial networks}

\author[1,*]{Martin Keller-Ressel}
\author[1]{Stephanie Nargang}
\affil[1]{TU Dresden, Institute for Mathematical Stochastics, Dresden, 01062, Germany}

\affil[*]{martin.keller-ressel@tu-dresden.de}

\begin{abstract}Based on data from the European banking stress tests of 2014, 2016 and the transparency exercise of 2018 we demonstrate for the first time that the latent geometry of financial networks can be well-represented by geometry of negative curvature, i.e., by hyperbolic geometry. This allows us to connect the network structure to the popularity-vs-similarity model of Papdopoulos et al., which is based on the Poincar\'e disc model of hyperbolic geometry. We show that the latent dimensions of `popularity' and `similarity' in this model are strongly associated to systemic importance and to geographic subdivisions of the banking system. In a longitudinal analysis over the time span from 2014 to 2018 we find that the systemic importance of individual banks has remained rather stable, while the peripheral community structure exhibits more (but still moderate) variability.
\end{abstract}
\begin{document}

\flushbottom
\maketitle
%
%
\thispagestyle{empty}

\section*{Introduction}
Network models based on hyperbolic geometry have been successful in explaining the structural features of informational\cite{shavitt2004curvature}, social\cite{muscoloni2017machine} and biological networks\cite{alanis2016manifold}. Such models provide a mathematical framework to resolve the conflicting paradigms of preferential attachment (attraction to \emph{popular} nodes) and community effects (attraction to \emph{similar} nodes) in networks. \cite{papadopoulos2012popularity, papadopoulos2015network, barabasi2012luck}\\
Just as the geometric structure of a social network determines the diffusion of news, rumors or infective diseases between individuals \cite{brockmann2013hidden}, the geometric structure of a financial network influences the diffusion of financial stress between financial institutions, such as banks.\cite{cont2010network, battiston2012liaisons, roukny2013default} Indeed, the lack of understanding for risks originating from the systemic interaction of financial institutions has been identified as a major contributing factor to the global financial crisis of 2008.\cite{french2010squam} While many recent studies have analysed the mechanisms of financial contagion in theoretical or simulation-based settings, less attention has been payed to the structural and geometric characteristics of real financial networks. In particular, it has remained an open question, whether the paradigm of hyperbolic structure applies to financial and economic networks and what such a structure implies for financial contagion processes.\\
Here, we consider financial networks inferred from bank balance sheet data, as collected and made available by the European Banking Authority (EBA) within the European banking stress test and transparency exercises of 2014, 2016 and 2018.\cite{EBA2016,EBA2018} We show that these networks can be embedded into low-dimensional hyperbolic space with considerably smaller distortion than into Euclidean space, suggesting that the paradigm of latent hyperbolic geometry also applies to financial networks. 
Furthermore -- following Papadopoulos et al.{}\cite{papadopoulos2012popularity} -- we decompose the embedded hyperbolic coordinates into the dimensions of \emph{popularity} and \emph{similarity} and demonstrate that these dimensions align with \emph{systemic importance} and membership in \emph{regional banking clusters} respectively. Finally, the longitudinal structure of the data allows us to track changes in these dimensions over time, i.e., to track the stability of systemic importance and of the peripheral community structure over time. 

\section*{Results}

\subsection*{Inference of Financial Networks}
Contagion in financial networks is a complex process, which can take place through several parallel (and potentially interacting) mechanisms and channels. \cite{caccioli2015overlapping} These mechanisms include direct bank-to-bank liabilities\cite{eisenberg2001systemic}, bank runs\cite{brown2017understanding}, and market-mediated contagion through asset sales \cite{shleifer1992liquidation,caccioli2015overlapping,glasserman2015likely,cont2017fire} (`fire-sale contagion'); see also \cite[p.21ff]{french2010squam}. Here, we focus on the channel of fire-sale contagion, which has been singled out -- both in simulation\cite{glasserman2015likely} and in empirical studies\cite{shleifer1992liquidation} --  as a key mechanism of financial contagion. Moreover, the propensity of fire-sale contagion can be quantified from available balance sheet data, using liquidity-weighted portfolio overlap (LWPO)\cite{cont2016fire, cont2017fire} as an indicator (see Methods for details).

Our inference of financial networks follows a two-stage mechanism: First, we construct a weighted bipartite network in which banks $B = (b_1, \dotsc, b_n)$ are linked to a common pool of assets $A = (a_1, \dotsc, a_m)$, which consist of sovereign bonds classified by issuing country and by different levels of maturity. In the second step we perform a one-mode projection of this network on the node set $B$, using the LWPO of two banks $b_i, b_j \in B$ to determine the weight $w_{ij}$ of the link between the corresponding nodes. For any of the years $y \in \{2014, 2016, 2018\}$, the result is an undirected, weighted network $N_y$ of banks, in which two banks are connected if and only if they hold common assets. The link weight $w_{ij}$, normalized to $[0,1]$, represents the susceptibility of two banks $b_i, b_j$ to financial contagion, quantified by their LWPO. The inferred networks are very dense, i.e., almost all pairs of banks hold \emph{some} common assets. However, most of the connections have very small weights, and the networks are dominated by a `sparse backbone' of a few strong connections, which represent the dominant channels of potential contagion of financial distress; see Figure~\ref{fig:comparison}.A. 

\subsection*{Latent network geometry}
\begin{wrapfigure}{r}{0.5\textwidth}
\includegraphics[width=0.48\textwidth]{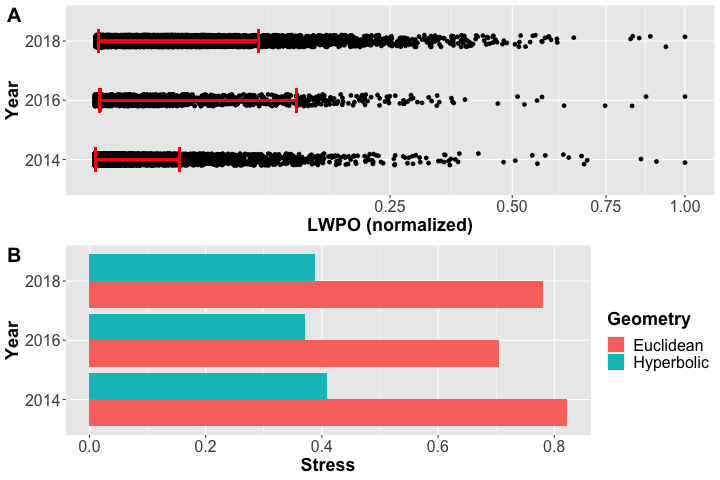}
{\footnotesize
\caption{\textit{Panel A}: Edge weight distribution in the EBA Networks of years 2014, 2016 and 2018. The inter-decile range of the (highly skewed) distributions is indicated in red.\\
\textit{Panel B}: Stress of network embedding into Euclidean vs. Hyperbolic geometry.  Lower values of stress indicate better goodness-of-fit.}\label{fig:comparison}}
\end{wrapfigure}
\begin{figure}[hp]
\centering
\includegraphics[width=0.66\linewidth]{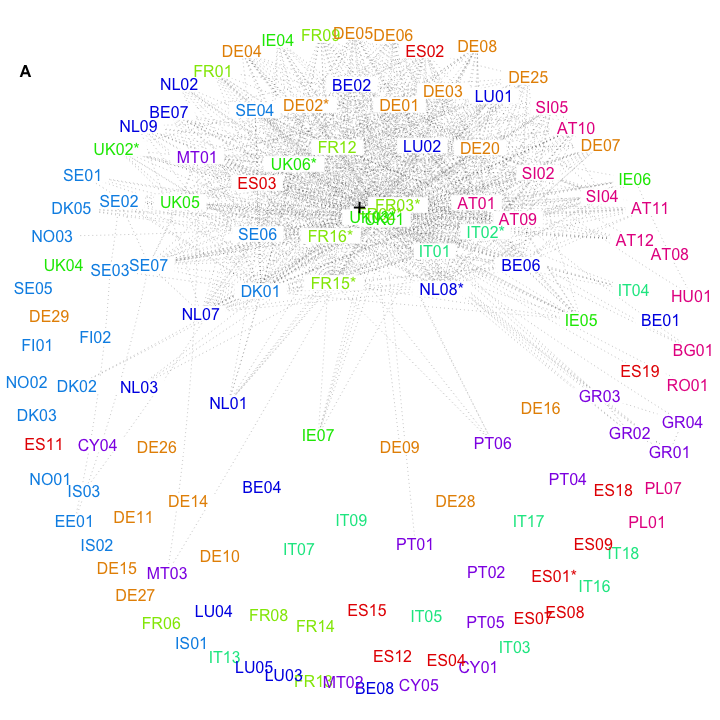}
\includegraphics[width=0.33\linewidth]{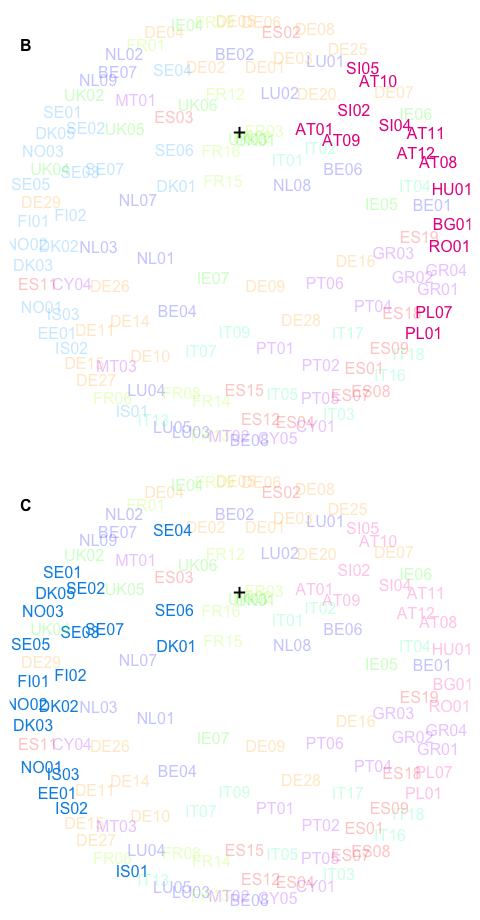}
\caption{Hyperbolic Embedding of the EBA Financial Network of 2018. Nodes are labelled by country and bank ID and coloured according to region (see Table~\ref{tab:banks_full} for full names). Panel A shows the full network including the strongest links (top decile), i.e., the connections with the largest liquidity-weighted portfolio overlap. Banks labelled as systemically important by the Financial Stability Board (G-SIBs) are indicated by asterisks. The black cross marks the capital-weighted hyperbolic center of the banking network. In panels B and C the Central/Eastern and the Nordic regional groups are highlighted to illustrate regional clustering.}
\label{fig:EBA2018}
\end{figure}

Our first objective was to uncover the latent geometric network structure and to evaluate the suitability of a hyperbolic network model. (See Methods for background on hyperbolic geometry.) To this end, we calculated stress-minimizing embeddings of the financial networks $N_{2014}, N_{2016}$ and $N_{2018}$ into two-dimensional Euclidean space $\mathbb{E}_2$ and hyperbolic space $\mathbb{H}_2$. These methods correspond to classic multidimensional scaling\cite{kruskal1978multidimensional} in the Euclidean case and to the \texttt{hydra+} embedding method\cite{chowdhary2017improved, keller2019hydra} in the hyperbolic case. The residual stress can be used as a goodness-of-fit measure between geometric model and true network topology. As shown in Figure~\ref{fig:comparison}.B the residual embedding stress from the hyperbolic model is substantially smaller -- consistently over all three years of observation -- than from the Euclidean model. This indicates that the latent geometry of the observed financial networks is much better represented by negatively curved (hyperbolic) rather than flat (Euclidean) geometry. It is also evidence of a high degree of hierarchical organization\cite{krioukov2010hyperbolic} in the financial networks considered. Furthermore, as a result of the embedding we obtain for each bank node $b_i$ latent coordinates $(r_i, \theta_i)$ in the Poincar\'e disc model of hyperbolic space (see Methods), which allows us to connect the network embedding to the popularity-vs-similarity model of Papadopoulos et al.\cite{papadopoulos2012popularity} The hyperbolic embedding of the full banking network of 2018 is shown in Figure~\ref{fig:EBA2018}. The embedded network shows a clear core-periphery structure, in line with previous studies of financial networks.\cite{boss2004network,langfield2014mapping} A deeper analysis of this structure is the subject of the following section.

\subsection*{Structural Analysis}
The popularity-vs-similarity model of Papadopoulos et al.\cite{papadopoulos2012popularity} offers a direct interpretation of the latent hyperbolic network coordinates in the Poincar\'e disc in terms of their \emph{popularity} dimension (the radial coordinate $r$) and the \emph{similarity} dimension (the angular coordinate $\theta$). In the context of financial networks, we hypothesized  that the popularity dimension of a given bank aligns with its systemic importance, and that its similarity dimension is associated with sub-sectors of the banking system, e.g., along geographic and regional divisions. Due to the asymmetric distribution of banks within the Poincar\'e disc (Figure~\ref{fig:EBA2018}), we slightly adapt the model Papadopoulos et al.\cite{papadopoulos2012popularity} and calculate the geodesic polar coordinates $(r'_i,\theta'_i)$ with respect to the network center-of-weight, rather than the center of the Poincar\'e disc; see Methods for details.\par
\vspace{0.4em}
\begin{wraptable}[14]{l}{0.6\textwidth}
{\footnotesize
\begin{center}
\begin{tabular}{cccc}
\toprule
Rank & 2014 & 2016 & 2018\\
\midrule
1 & Nordea *& BNP Paribas *& Groupe BPCE * \\
2 & Royal Bank of Scotland *& UniCredit *& Barclays *\\
3 & Barclays * & ING Groep *& Royal Bank of Scotland \\
4 & Intesa Sanpaolo & Deutsche Bank *& Groupe Cr\'edit Agricole *\\
5 & UniCredit * &Intesa Sanpaolo & BNP Paribas *\\
\bottomrule
\end{tabular}
\end{center}
}
\caption{For each year the five banks with the highest hyperbolic centrality (i.e. smallest $r'$ coordinate) are listed. Asterisks denote banks that are considered globally systemic relevant institutions (G-SIBs)}
\label{tab:top}
\end{wraptable}
To test the first hypothesis -- the association between radial coordinate $r'$ and systemic importance -- we labelled a bank as \emph{systemically important} in a given year, whenever it was included in the contemporaneous list of global systemically important banks (G-SIBs) as published by the Financial Stability Board.\cite{GSIB2014, GSIB2016, GSIB2018}; see also Table~\ref{tab:banks_full}. Using a Wilcoxon–Mann–Whitney test, we find a significant association between radial rank and systemic importance in all years ($P_{2014} < .0001$, $P_{2016} < .0001$, $P_{2018} = .0038$). In Table \ref{tab:top} we report the five top-ranked banks (most central in terms of $r'$) for each year.\\

To test the second hypothesis -- the association between similarity dimension $\theta'$ and regional banking sub-sectors --  we assigned banks to the following nine regional groups:
\begin{quote} 
Spain (ES), Germany (DE), France (FR), Italy (IT), UK and Ireland (UK/IE), Nordic Region (EE/NO/SE/DK/FI/IS), Benelux Region (BE/NE/LU), Southern/Mediterranean (GR/CY/MT/PT), Central and Eastern Europe (AT/BG/HU/LV/RO/SI).
\end{quote}
These regions are reasonably balanced in terms of the number of banks included in the EBA panel. Using ANOVA for circular data\cite[Sec.~7.4]{mardia2009directional} we find a highly significant association between the angular coordinate $\theta'$ and the regional group in all three years considered ($P < .0001$ in all years). This indicates that the peripheral community structure (away from the network core) of the EBA financial network is indeed strongly aligned with geographic and regional divisions in Europe. We have highlighted two different regional groups in Figure~\ref{fig:EBA2018}.B and \ref{fig:EBA2018}.C to illustrate the association between angular coordinate and regional structure. 


\subsection*{Network structure over time}
The longitudinal structure of the data set allows us to track changes in the network structure over the whole time span of observations from 2014 to 2018. Note, however, that the samples of banks included by the EBA vary substantially in size and -- even when restricted to the smallest sample -- are not completely overlapping; see Table~\ref{tab:sample}. Nevertheless, the goodness-of-fit of the hyperbolic model (reported in Figure~\ref{fig:comparison}.B) is surprisingly stable over all years. This suggests that the hyperbolic model does indeed capture intrinsic qualities of the network, rather than relying on transitory structural artefacts.\\
 \begin{wrapfigure}{r}{0.6\textwidth}
\includegraphics[width=0.58\textwidth]{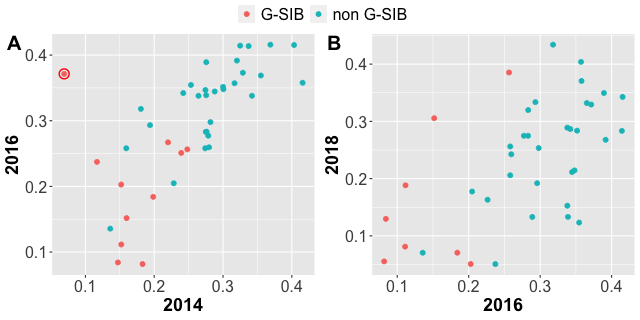}
\caption{Changes in radial coordinate $r'$ (low values indicate high centrality) between 2014 and 2016 (A) and 2016 and 2018 (B). Banks considered systemically relevant (G-SIBs) at the end of the time period are marked in red. Nordea bank is circled in the panel A; see text for background.\label{fig:longitudinal}}
\end{wrapfigure}
We proceed to analyze the temporal changes in the latent radial coordinate $r'$ and angular coordinate $\theta'$, corresponding to changes in systemic importance and community structure. Note that the small sample of banks included in the 2016 stress test restricts the number of banks that are included in this longitudinal analysis, cf. Table~\ref{tab:sample}. The scatter plots in Figure~\ref{fig:longitudinal} and the corresponding Pearson's correlations of $.678$ between 2014 and 2016 $(P < .0001)$ and $.569$ between 2016 and 2018 $(P = .0001)$ show a clear positive association between hyperbolic centrality in successive snap shots of the financial networks. In panel $A$ of Figure~\ref{fig:longitudinal}, Nordea bank can be identified as a clear outlier, moving from a very central position in 2014 to a peripheral position in 2016. Interestingly, Nordea was one of just two banks (together with Royal Bank of Scotland) which were removed from the list of G-SIBs in the subsequent update in 2018 due to decreasing systemic importance.\cite{GSIB2018}\\

For the angular coordinate, we account for the circular nature of the variable and compute the \emph{circular correlation}~\cite{mardia2009directional} of the angular coordinates between successive years. A moderate association between successive years can be observed at circular correlation values of $0.211$ between 2014 and 2016 $(P = .1877)$ and $0.225$ between 2016 and 2018 ($P = 0.1385$). 
\section*{Discussion}
Based on data from the EBA stress tests of 2014, 2016 and the transparency exercise of 2018, we have presented strong evidence that the latent geometry of financial networks can be well-represented by geometry of negative curvature, i.e., by hyperbolic geometry. Calculating stress-minimizing embeddings into the Poicar\'e disc model of hyperbolic geometry has allowed us to visualize this geometric structure and to connect it to the popularity-vs-similarity model of Papdopoulos et al.\cite{papadopoulos2012popularity} We find that the radial coordinate (\emph{`popularity'}) is strongly associated with systemic importance (as assessed by the Financial Stability Board) and the angular coordinate (\emph{`similarity'}) with geographic and regional subdivisions. A longitudinal analysis shows that -- in the observation period from 2014 to 2018 -- systemic importance of banks within the European banking network has stayed rather stable and has been predominated by only gradual changes. The peripheral community structure has been more variable, but has remained strongly determined by geographical divisions in all years considered.\\
In future research we plan to study the interplay between hyperbolic network geometry and the dynamics of contagion processes. We are confident, that the empirical analysis of latent network geometry in this paper can provide the basis for new analytic models for the diffusion of financial stress in a banking network with hyperbolic structure.

\section*{Methods}
\subsection*{Data Preparation and Inference of Financial Networks}


The financial networks were extracted from three different publicly available data sets stemming from the stress tests (in 2014 and 2016) and the EU-wide transparency exercise (in 2018) of the European Banking Authority (EBA).\cite{EBA2016,EBA2018} The data sets contain detailed balance sheet information from all European banks (EU + Norway) included in the stress test/transparency exercise of the EBA in the respective year. From these data sets we extracted the portfolio values of all sovereign bonds held by the banks, split by issuing country (38 countries) and three levels of maturity (short: 0M-3M, medium: 3M-2Y, long: 2Y-10Y+), resulting in $m = 38 \times 3 = 114$ different asset classes. 

\begin{wraptable}{r}{0.5\textwidth}
{\footnotesize
\begin{center}
\begin{tabular}{lccc}
\toprule
 & 2014 & 2016 & 2018\\
\midrule
number of banks ($n)$ & 119 & 51 & 128\\
of which included in the subseq.{} year & 43 & 41 & \\
\bottomrule
\end{tabular}
\end{center}
\caption{Sample sizes of EBA data sets\label{tab:sample}}}
\end{wraptable}
For each year, this data was stored as the weighted adjacency matrix $P$ (`portfolio matrix') of a bipartite network. The $n$ rows of $P$  correspond to the banks in the sample, the $m$ columns to the different asset classes, and the element $P_{ik}$ to the portfolio value (in EUR) of asset $k$ in the balance sheet of bank $i$. To perform a one-mode projection of this bipartite network, we followed Cont and Wagalath\cite{cont2013running, cont2016fire} as well as Cont and Schaanning\cite{cont2017fire}: We computed the liquidity-weighted portfolio overlap (LWPO) of bank $i$ and bank $j$ as 
\begin{equation}\label{eq:LWPO}
 L_{ij} = \sum_{k=1}^m \frac{P_{ik}  P_{jk}}{d_k},
 \end{equation}
where $d_k$ is the market depth for asset $k$.\cite{cont2017fire} The LWPO measures the impact of a sudden liquidation of the portfolio of bank $i$ on the portfolio value of bank $j$ and vice versa. Hence, it quantifies the risk of fire-sale contagion between the banks in a financial stress scenario. The market depth of asset $k$ was estimated from $P$ as its total volume held by all banks in the sample, i.e., as $d_k = \sum_{i=1}^n P_{ik}$. Writing $D$ for the diagonal matrix of market depths,  \eqref{eq:LWPO} can be succinctly written as matrix product $L = P D^{-1} P^\top$. Finally, we set the link weight $w_{ij}$ between bank $b_i$ and $b_j$ in the one-mode projection $N$ of the banking network equal to the normalized LWPO between banks $b_i$ and $b_j$, i.e., $w_{ij} := L_{ij} / \max_{i,j}L_{ij}$

\subsection*{Background on hyperbolic geometry} 
\subsubsection*{The hyperboloid model}
Hyperbolic geometry can be characterized as the geometry of a space of constant \emph{negative} curvature, while the more familiar Euclidean geometry is the geometry of a flat space, i.e. a space of zero curvature. In the \emph{hyperboloid model} of hyperbolic geometry\cite{ratcliffe1994foundations, cannon1997hyperbolic}, $d$-dimensional hyperbolic space $\mathbb{H}_d$ is defined as the hyperboloid
\[\mathbb{H}_d = \set{x \in \mathbb{R}^{d+1}: x_0^2 - x_1^2 - \dotsm - x_d^2 = 1, \; x_0 > 0} \quad \text{equipped with distance} \quad 
\dist_H(x,y) = \arcosh\left(x_0 y_0 - x_1 y_1 - \dotsm  - x_d y_d\right).
\]
In fact, $\mathbb{H}_d$ endowed with the Riemannian metric tensor
$ds^2 = dx_0^2 - dx_1^2 - \dotsm - dx_d^2$
is a Riemannian manifold and $\dist_H(x,y)$ is the corresponding Riemannian distance.\cite{ratcliffe1994foundations, cannon1997hyperbolic} The sectional curvature of this manifold is constant and equal to $-1$. Thus, $\mathbb{H}_d$ is indeed a model of geometry of constant negative curvature.
\subsubsection*{The Poincar\'e disc model}
While the hyperboloid model is convenient for computations, a more preferable (and popular) model for visualizations in dimension $d=2$ is the \emph{Poincar\'e disc model}\cite{ratcliffe1994foundations}, which also forms the basis of the popularity-vs-similarity model of Papadopoulos et al.\cite{papadopoulos2012popularity}. To obtain the Poincar\'e disc model, the hyperboloid $\mathbb{H}_2$ is mapped to the open unit disc (`Poincar\'e disc') $\mathbb{D} = \set{z \in \mathbb{R}^2: z_1^2 + z_2^2 < 1}$, parameterized by polar coordinates as $z_1 = r \cos \theta$, $z_2 = r \sin \theta$, using the 
\emph{stereographic projection} (cf. \cite[\S4.2]{ratcliffe1994foundations}) 
\begin{align}\label{eq:conversion_r}
r = \sqrt{\frac{x_0 -1}{x_0 + 1}},  \qquad \theta = \atan2(x_2,x_1), \qquad x = (x_0,x_1,x_2) \in \mathbb{H}_2,
\end{align}
where $\atan2$ is the quadrant-preserving arctangent.\footnote{The quadrant-preserving arctangent $\atan2(x_2,x_1)$, well-defined unless $x_1 = x_2 = 0$, returns the unique angle $\theta \in [0,2\pi)$ which solves $\tan \theta = x_2/x_1$ and points to the same quadrant as $(x_1,x_2)$. It is commonly implemented in scientific computing environments (e.g. in MATLAB or \textsf{R}) as \texttt{atan2}.} In the Poincar\'e disc model, the hyperbolic distance becomes
\[\dist_B((r_1,\theta_1),(r_2,\theta_2)) = \arcosh\left(1 + 2\frac{r_1^2 + r_2^2 + 2r_1 r_2 \cos(\theta_1 - \theta_2)}{(1 - r_1^2)(1 - r_2^2)}\right)\]
and geodesic lines are represented by arcs of (Euclidean) circles intersected with $\mathbb{D}$.
\subsection*{Hyperbolic Embedding and Centering}
\subsubsection*{Embedding into Hyperbolic Space}
Network embedding methods aim to find -- for each network node $b_i$  -- latent coordinates $x^i$ in a geometric model space $G$, such that the geodesic distance between $x^i$ and $x^j$ in $G$ matches -- as closely as possible -- a given dissimilarity measure $\dist_{ij}$ between nodes $b_i$ and $b_j$. Stress-minimizing embedding methods aim to minimize the stress functional 
\begin{equation}\label{eq:stress}
\text{Stress}(x^1, \dotsc, x^n) = \sqrt{\frac{1}{n(n-1)}\sum_{i,j} \left(\dist_{ij}^\text{network} - \dist^\text{geom}_G(x^i, x^j)\right)^2},
\end{equation}
which measures the root mean square error between given network dissimilarities and the corresponding distances in the model space. For Euclidean geometry, this method is well-known as multidimensional scaling\cite{kruskal1978multidimensional,borg2003modern}, or  -- using a weighted stress functional -- as Sammon mapping\cite{sammon1969nonlinear}. For hyperbolic space, i.e., when $\dist_G^\text{geom} = \dist_H$, several optimization methods for \eqref{eq:stress} have been proposed\cite{zhao2011fast, chowdhary2017improved,keller2019hydra}. We use the \texttt{hydra+} method implemented in the package \texttt{hydra} for the statistical computing environment \textsf{R}\cite{r2019}. 
\subsubsection*{Hyperbolic Centering}
For a point cloud $x^1, \dots, x^n$ in $\mathbb{H}_d$ and non-negative weights $w_1, \dotsc, w_n$ summing to one, the \emph{hyperbolic mean}\cite{mardia2009directional} or \emph{hyperbolic center of weight}\cite{galperin1993concept} can be determined as follows: Calculate the weighted Euclidean mean $\bar x = \sum w_i x^i$, and its `resultant length' $\rho = \sqrt{(\bar x_0)^2 - (\bar x_1)^2 - \dotsm  - (\bar x_d)^2}$, which is a measure of dispersion for the point cloud. The hyperbolic center $c$ is then determined as $c = \bar x / \rho$ and is again an element of $\mathbb{H}_d$. The point cloud can be centered at $c$ by transforming each points as $(x^i)' = T_{-c} x^i$, where $T_{c}$ is the hyperbolic translation matrix (`Lorentz boost') 
\[T_{c} = \begin{pmatrix} c_0 & \bar c^\top \\ \bar c & \sqrt{I_d + \bar c \bar c^\top}\end{pmatrix} \qquad \text{with} \qquad c=(c_0,\bar c) = (c_0, c_1, \dotsc, c_d).\]
In dimension $d = 2$, the stereographic projection \eqref{eq:conversion_r} may then be applied to convert the centered coordinates $(x^i)'$ to centered polar coordinates $(r'_i,  \theta'_i)$ in the Poincar\'e disc.

\subsubsection*{Application to Financial Networks}
The described methods were applied to the financial networks inferred from the EBA data as follows: We converted the similarity weights $w_{ij}$ (normalized LWPO) to dissimilarities $\dist_{ij} = 1 - w_{ij}$. We embedded these similarities by minimizing the stress functional \eqref{eq:stress}, using the R-package \texttt{hydra}. For the resulting network embeddings, we calculated the capital-weighted network center $c$ as the weighted hyperbolic mean with weights $w_i$ proportional to the total capital $\sum_{k=1}^m P_{ik}$ of bank $i$ invested in all assets $a_1, \dots a_m$. After centering at the hyperbolic center $c$, we calculated the coordinates $(r'_i, \theta'_i)$ by the stereographic projection \eqref{eq:conversion_r}.
 
\subsection*{Data Availability Statement}
The data analysed during the current study are available from the website of the European Banking Authority at \url{https://www.eba.europa.eu/risk-analysis-and-data/eu-wide-stress-testing} and \url{https://eba.europa.eu/risk-analysis-and-data/eu-wide-transparency-exercise/2018}.

\bibliography{sample}

\begin{thebibliography}{10}
\urlstyle{rm}
\expandafter\ifx\csname url\endcsname\relax
  \def\url#1{\texttt{#1}}\fi
\expandafter\ifx\csname urlprefix\endcsname\relax\def\urlprefix{URL }\fi
\expandafter\ifx\csname doiprefix\endcsname\relax\def\doiprefix{DOI: }\fi
\providecommand{\bibinfo}[2]{#2}
\providecommand{\eprint}[2][]{\url{#2}}

\bibitem{shavitt2004curvature}
\bibinfo{author}{Shavitt, Y.} \& \bibinfo{author}{Tankel, T.}
\newblock \bibinfo{title}{On the curvature of the internet and its usage for
  overlay construction and distance estimation}.
\newblock In \emph{\bibinfo{booktitle}{IEEE INFOCOM 2004}},
  vol.~\bibinfo{volume}{1} (\bibinfo{organization}{IEEE},
  \bibinfo{year}{2004}).

\bibitem{muscoloni2017machine}
\bibinfo{author}{Muscoloni, A.}, \bibinfo{author}{Thomas, J.~M.},
  \bibinfo{author}{Ciucci, S.}, \bibinfo{author}{Bianconi, G.} \&
  \bibinfo{author}{Cannistraci, C.~V.}
\newblock \bibinfo{journal}{\bibinfo{title}{Machine learning meets complex
  networks via coalescent embedding in the hyperbolic space}}.
\newblock {\emph{\JournalTitle{Nature communications}}}
  \textbf{\bibinfo{volume}{8}}, \bibinfo{pages}{1--19} (\bibinfo{year}{2017}).

\bibitem{alanis2016manifold}
\bibinfo{author}{Alanis-Lobato, G.}, \bibinfo{author}{Mier, P.} \&
  \bibinfo{author}{Andrade-Navarro, M.~A.}
\newblock \bibinfo{journal}{\bibinfo{title}{Manifold learning and maximum
  likelihood estimation for hyperbolic network embedding}}.
\newblock {\emph{\JournalTitle{Applied network science}}}
  \textbf{\bibinfo{volume}{1}}, \bibinfo{pages}{1--14} (\bibinfo{year}{2016}).

\bibitem{papadopoulos2012popularity}
\bibinfo{author}{Papadopoulos, F.}, \bibinfo{author}{Kitsak, M.},
  \bibinfo{author}{Serrano, M.~{\'A}.}, \bibinfo{author}{Bogun{\'a}, M.} \&
  \bibinfo{author}{Krioukov, D.}
\newblock \bibinfo{journal}{\bibinfo{title}{Popularity versus similarity in
  growing networks}}.
\newblock {\emph{\JournalTitle{Nature}}} \textbf{\bibinfo{volume}{489}},
  \bibinfo{pages}{537} (\bibinfo{year}{2012}).

\bibitem{papadopoulos2015network}
\bibinfo{author}{Papadopoulos, F.}, \bibinfo{author}{Psomas, C.} \&
  \bibinfo{author}{Krioukov, D.}
\newblock \bibinfo{journal}{\bibinfo{title}{Network mapping by replaying
  hyperbolic growth}}.
\newblock {\emph{\JournalTitle{IEEE/ACM Transactions on Networking (TON)}}}
  \textbf{\bibinfo{volume}{23}}, \bibinfo{pages}{198--211}
  (\bibinfo{year}{2015}).

\bibitem{barabasi2012luck}
\bibinfo{author}{Barabasi, A.-L.}
\newblock \bibinfo{journal}{\bibinfo{title}{Luck or reason}}.
\newblock {\emph{\JournalTitle{Nature}}} \textbf{\bibinfo{volume}{486}},
  \bibinfo{pages}{507--509} (\bibinfo{year}{2012}).

\bibitem{brockmann2013hidden}
\bibinfo{author}{Brockmann, D.} \& \bibinfo{author}{Helbing, D.}
\newblock \bibinfo{journal}{\bibinfo{title}{The hidden geometry of complex,
  network-driven contagion phenomena}}.
\newblock {\emph{\JournalTitle{Science}}} \textbf{\bibinfo{volume}{342}},
  \bibinfo{pages}{1337--1342} (\bibinfo{year}{2013}).

\bibitem{cont2010network}
\bibinfo{author}{Cont, R.}, \bibinfo{author}{Moussa, A.} \&
  \bibinfo{author}{Santos, E.~B.}
\newblock \bibinfo{title}{Network structure and systemic risk in banking
  systems}.
\newblock In \bibinfo{editor}{Jean-Pierre~Fouque, J. A.~L.} (ed.)
  \emph{\bibinfo{booktitle}{Network Structure and Systemic Risk in Banking
  Systems}} (\bibinfo{publisher}{Cambridge University Press},
  \bibinfo{year}{2010}).

\bibitem{battiston2012liaisons}
\bibinfo{author}{Battiston, S.}, \bibinfo{author}{Gatti, D.~D.},
  \bibinfo{author}{Gallegati, M.}, \bibinfo{author}{Greenwald, B.} \&
  \bibinfo{author}{Stiglitz, J.~E.}
\newblock \bibinfo{journal}{\bibinfo{title}{Liaisons dangereuses: Increasing
  connectivity, risk sharing, and systemic risk}}.
\newblock {\emph{\JournalTitle{Journal of economic dynamics and control}}}
  \textbf{\bibinfo{volume}{36}}, \bibinfo{pages}{1121--1141}
  (\bibinfo{year}{2012}).

\bibitem{roukny2013default}
\bibinfo{author}{Roukny, T.}, \bibinfo{author}{Bersini, H.},
  \bibinfo{author}{Pirotte, H.}, \bibinfo{author}{Caldarelli, G.} \&
  \bibinfo{author}{Battiston, S.}
\newblock \bibinfo{journal}{\bibinfo{title}{Default cascades in complex
  networks: Topology and systemic risk}}.
\newblock {\emph{\JournalTitle{Scientific reports}}}
  \textbf{\bibinfo{volume}{3}}, \bibinfo{pages}{2759} (\bibinfo{year}{2013}).

\bibitem{french2010squam}
\bibinfo{author}{French, K.} \emph{et~al.}
\newblock \bibinfo{journal}{\bibinfo{title}{The {S}quam {L}ake report: fixing
  the financial system}}.
\newblock {\emph{\JournalTitle{Journal of Applied Corporate Finance}}}
  \textbf{\bibinfo{volume}{22}}, \bibinfo{pages}{8--21} (\bibinfo{year}{2010}).

\bibitem{EBA2016}
\bibinfo{author}{{European Banking Authority}}.
\newblock \bibinfo{title}{{EU}-wide stress testing}.
\newblock
  \bibinfo{howpublished}{\url{https://www.eba.europa.eu/risk-analysis-and-data/eu-wide-stress-testing}}.

\bibitem{EBA2018}
\bibinfo{author}{{European Banking Authority}}.
\newblock \bibinfo{title}{{EU}-wide transparency exercise}.
\newblock
  \bibinfo{howpublished}{\url{https://eba.europa.eu/risk-analysis-and-data/eu-wide-transparency-exercise/2018}}.

\bibitem{caccioli2015overlapping}
\bibinfo{author}{Caccioli, F.}, \bibinfo{author}{Farmer, J.~D.},
  \bibinfo{author}{Foti, N.} \& \bibinfo{author}{Rockmore, D.}
\newblock \bibinfo{journal}{\bibinfo{title}{{Overlapping portfolios, contagion,
  and financial stability}}}.
\newblock {\emph{\JournalTitle{Journal of Economic Dynamics and Control}}}
  \textbf{\bibinfo{volume}{51}}, \bibinfo{pages}{50--63},
  \doiprefix\url{10.1016/j.jedc.2014.09.041} (\bibinfo{year}{2015}).

\bibitem{eisenberg2001systemic}
\bibinfo{author}{Eisenberg, L.} \& \bibinfo{author}{Noe, T.~H.}
\newblock \bibinfo{journal}{\bibinfo{title}{Systemic risk in financial
  systems}}.
\newblock {\emph{\JournalTitle{Management Science}}}
  \textbf{\bibinfo{volume}{47}}, \bibinfo{pages}{236--249}
  (\bibinfo{year}{2001}).

\bibitem{brown2017understanding}
\bibinfo{author}{Brown, M.}, \bibinfo{author}{Trautmann, S.~T.} \&
  \bibinfo{author}{Vlahu, R.}
\newblock \bibinfo{journal}{\bibinfo{title}{Understanding bank-run contagion}}.
\newblock {\emph{\JournalTitle{Management Science}}}
  \textbf{\bibinfo{volume}{63}}, \bibinfo{pages}{2272--2282}
  (\bibinfo{year}{2017}).

\bibitem{shleifer1992liquidation}
\bibinfo{author}{Shleifer, A.} \& \bibinfo{author}{Vishny, R.~W.}
\newblock \bibinfo{journal}{\bibinfo{title}{{Liquidation Values and Debt
  Capacity: A Market Equilibrium Approach}}}.
\newblock {\emph{\JournalTitle{The Journal of Finance}}}
  \textbf{\bibinfo{volume}{47}}, \bibinfo{pages}{1343--1366},
  \doiprefix\url{10.1111/j.1540-6261.1992.tb04661.x} (\bibinfo{year}{1992}).

\bibitem{glasserman2015likely}
\bibinfo{author}{Glasserman, P.} \& \bibinfo{author}{Young, H.~P.}
\newblock \bibinfo{journal}{\bibinfo{title}{{How likely is contagion in
  financial networks?}}}
\newblock {\emph{\JournalTitle{Journal of Banking and Finance}}}
  \textbf{\bibinfo{volume}{50}}, \bibinfo{pages}{383--399},
  \doiprefix\url{10.1016/j.jbankfin.2014.02.006} (\bibinfo{year}{2015}).

\bibitem{cont2017fire}
\bibinfo{author}{Cont, R.} \& \bibinfo{author}{Schaanning, E.}
\newblock \bibinfo{title}{Fire sales, indirect contagion and systemic stress
  testing} (\bibinfo{year}{2017}).
\newblock \bibinfo{note}{Norges Bank Working Paper 02/2017}.

\bibitem{cont2016fire}
\bibinfo{author}{Cont, R.} \& \bibinfo{author}{Wagalath, L.}
\newblock \bibinfo{journal}{\bibinfo{title}{Fire sales forensics: measuring
  endogenous risk}}.
\newblock {\emph{\JournalTitle{Mathematical Finance}}}
  \textbf{\bibinfo{volume}{26}}, \bibinfo{pages}{835--866}
  (\bibinfo{year}{2016}).

\bibitem{kruskal1978multidimensional}
\bibinfo{author}{Kruskal, J.~B.} \& \bibinfo{author}{Wish, M.}
\newblock \emph{\bibinfo{title}{Multidimensional scaling}},
  vol.~\bibinfo{volume}{11} (\bibinfo{publisher}{Sage}, \bibinfo{year}{1978}).

\bibitem{chowdhary2017improved}
\bibinfo{author}{Chowdhary, K.} \& \bibinfo{author}{Kolda, T.~G.}
\newblock \bibinfo{journal}{\bibinfo{title}{An improved hyperbolic embedding
  algorithm}}.
\newblock {\emph{\JournalTitle{Journal of Complex Networks}}}
  \textbf{\bibinfo{volume}{6}}, \bibinfo{pages}{321--341}
  (\bibinfo{year}{2017}).

\bibitem{keller2019hydra}
\bibinfo{author}{Keller-Ressel, M.} \& \bibinfo{author}{Nargang, S.}
\newblock \bibinfo{journal}{\bibinfo{title}{Hydra: a method for
  strain-minimizing hyperbolic embedding of network-and distance-based data}}.
\newblock {\emph{\JournalTitle{Journal of Complex Networks}}}
  \textbf{\bibinfo{volume}{8}}, \bibinfo{pages}{cnaa002}
  (\bibinfo{year}{2020}).

\bibitem{krioukov2010hyperbolic}
\bibinfo{author}{Krioukov, D.}, \bibinfo{author}{Papadopoulos, F.},
  \bibinfo{author}{Kitsak, M.}, \bibinfo{author}{Vahdat, A.} \&
  \bibinfo{author}{Bogun{\'a}, M.}
\newblock \bibinfo{journal}{\bibinfo{title}{Hyperbolic geometry of complex
  networks}}.
\newblock {\emph{\JournalTitle{Physical Review E}}}
  \textbf{\bibinfo{volume}{82}}, \bibinfo{pages}{036106}
  (\bibinfo{year}{2010}).

\bibitem{boss2004network}
\bibinfo{author}{Boss, M.}, \bibinfo{author}{Elsinger, H.},
  \bibinfo{author}{Summer, M.} \& \bibinfo{author}{Thurner~4, S.}
\newblock \bibinfo{journal}{\bibinfo{title}{Network topology of the interbank
  market}}.
\newblock {\emph{\JournalTitle{Quantitative finance}}}
  \textbf{\bibinfo{volume}{4}}, \bibinfo{pages}{677--684}
  (\bibinfo{year}{2004}).

\bibitem{langfield2014mapping}
\bibinfo{author}{Langfield, S.}, \bibinfo{author}{Liu, Z.} \&
  \bibinfo{author}{Ota, T.}
\newblock \bibinfo{journal}{\bibinfo{title}{Mapping the {UK} interbank
  system}}.
\newblock {\emph{\JournalTitle{Journal of Banking \& Finance}}}
  \textbf{\bibinfo{volume}{45}}, \bibinfo{pages}{288--303}
  (\bibinfo{year}{2014}).

\bibitem{GSIB2014}
\bibinfo{author}{{Financial Stability Board}}.
\newblock \bibinfo{title}{2014 update of list of global systemically important
  banks ({G-SIB}s)}.
\newblock
  \bibinfo{howpublished}{\url{https://www.fsb.org/2014/11/2014-update-of-list-of-global-systemically-important-banks/}}.

\bibitem{GSIB2016}
\bibinfo{author}{{Financial Stability Board}}.
\newblock \bibinfo{title}{2016 list of global systemically important banks
  ({G-SIB}s)}.
\newblock
  \bibinfo{howpublished}{\url{https://www.fsb.org/2016/11/2016-list-of-global-systemically-important-banks-g-sibs/}}.

\bibitem{GSIB2018}
\bibinfo{author}{{Financial Stability Board}}.
\newblock \bibinfo{title}{2018 list of global systemically important banks
  ({G-SIB}s)}.
\newblock
  \bibinfo{howpublished}{\url{https://www.fsb.org/2018/11/2018-list-of-global-systemically-important-banks-g-sibs/}}.

\bibitem{mardia2009directional}
\bibinfo{author}{Mardia, K.~V.} \& \bibinfo{author}{Jupp, P.~E.}
\newblock \emph{\bibinfo{title}{Directional statistics}}
  (\bibinfo{publisher}{John Wiley \& Sons}, \bibinfo{year}{2009}).

\bibitem{cont2013running}
\bibinfo{author}{Cont, R.} \& \bibinfo{author}{Wagalath, L.}
\newblock \bibinfo{journal}{\bibinfo{title}{Running for the exit: distressed
  selling and endogenous correlation in financial markets}}.
\newblock {\emph{\JournalTitle{Mathematical Finance: An International Journal
  of Mathematics, Statistics and Financial Economics}}}
  \textbf{\bibinfo{volume}{23}}, \bibinfo{pages}{718--741}
  (\bibinfo{year}{2013}).

\bibitem{ratcliffe1994foundations}
\bibinfo{author}{Ratcliffe, J.~G.}
\newblock \emph{\bibinfo{title}{Foundations of hyperbolic manifolds}},
  vol.~\bibinfo{volume}{3} (\bibinfo{publisher}{Springer},
  \bibinfo{year}{1994}).

\bibitem{cannon1997hyperbolic}
\bibinfo{author}{Cannon, W.~J.}, \bibinfo{author}{Floyd, W.~J.},
  \bibinfo{author}{Kenyon, R.} \& \bibinfo{author}{Parry, W.~R.}
\newblock \bibinfo{title}{Hyperbolic geometry}.
\newblock In \bibinfo{editor}{{Silvio Levy}} (ed.)
  \emph{\bibinfo{booktitle}{Flavors of Geometry}}, \bibinfo{pages}{59--115}
  (\bibinfo{publisher}{{MSRI} Publications}, \bibinfo{year}{1997}),
  \bibinfo{edition}{31} edn.

\bibitem{borg2003modern}
\bibinfo{author}{Borg, I.} \& \bibinfo{author}{Groenen, P.}
\newblock \bibinfo{journal}{\bibinfo{title}{Modern multidimensional scaling:
  Theory and applications}}.
\newblock {\emph{\JournalTitle{Journal of Educational Measurement}}}
  \textbf{\bibinfo{volume}{40}}, \bibinfo{pages}{277--280}
  (\bibinfo{year}{2003}).

\bibitem{sammon1969nonlinear}
\bibinfo{author}{Sammon, J.~W.}
\newblock \bibinfo{journal}{\bibinfo{title}{A nonlinear mapping for data
  structure analysis}}.
\newblock {\emph{\JournalTitle{IEEE Transactions on computers}}}
  \textbf{\bibinfo{volume}{100}}, \bibinfo{pages}{401--409}
  (\bibinfo{year}{1969}).

\bibitem{zhao2011fast}
\bibinfo{author}{Zhao, X.}, \bibinfo{author}{Sala, A.}, \bibinfo{author}{Zheng,
  H.} \& \bibinfo{author}{Zhao, B.~Y.}
\newblock \bibinfo{journal}{\bibinfo{title}{Fast and scalable analysis of
  massive social graphs}}.
\newblock {\emph{\JournalTitle{arXiv preprint arXiv:1107.5114}}}
  (\bibinfo{year}{2011}).

\bibitem{r2019}
\bibinfo{author}{{R Core Team}}.
\newblock \emph{\bibinfo{title}{R: A Language and Environment for Statistical
  Computing}}.
\newblock \bibinfo{organization}{R Foundation for Statistical Computing},
  \bibinfo{address}{Vienna, Austria} (\bibinfo{year}{2019}).

\bibitem{galperin1993concept}
\bibinfo{author}{Galperin, G.}
\newblock \bibinfo{journal}{\bibinfo{title}{A concept of the mass center of a
  system of material points in the constant curvature spaces}}.
\newblock {\emph{\JournalTitle{Communications in Mathematical Physics}}}
  \textbf{\bibinfo{volume}{154}}, \bibinfo{pages}{63--84}
  (\bibinfo{year}{1993}).

\end{thebibliography}

\section*{Author contributions statement}
M.K.R. conceived the study,  S.N. prepared the data, M.K.R. and S.N. analysed the results.  All authors reviewed the manuscript. 

\section*{Additional information}
The author(s) declare no competing interests.

\begin{appendix}

\begin{table}[ht]
\centering
\footnotesize
\begin{tabular}{ll}
AT01 & Erste Group Bank AG  \\ 
AT08 & BAWAG Group AG  \\ 
AT09 & Raiffeisen Bank International AG  \\ 
AT10 & Raiffeisenbankengruppe Verbund eGen  \\ 
AT11 & Sberbank Europe AG  \\ 
AT12 & Volksbanken Verbund  \\ 
BE01 & Belfius Banque SA  \\ 
BE02 & Dexia NV  \\ 
BE04 & AXA Bank Europe SA  \\ 
BE06 & KBC Group NV  \\ 
BE07 & The Bank of New York Mellon SA/NV  \\ 
BE08 & Investar  \\ 
BG01 & First Investment Bank  \\ 
CY01 & Hellenic Bank Public Company Ltd  \\ 
CY04 & Bank of Cyprus Holdings Public Limited Company  \\ 
CY05 & RCB Bank Ltd   \\ 
DE01 & NRW.Bank  \\ 
DE02 & Deutsche Bank AG * \\ 
DE03 & Commerzbank AG  \\ 
DE04 & Landesbank Baden-W\"urttemberg  \\ 
DE05 & Bayerische Landesbank  \\ 
DE06 & Norddeutsche Landesbank-Girozentrale  \\ 
DE07 & Landesbank Hessen-Th\"uringen Girozentrale  \\ 
DE08 & DekaBank Deutsche Girozentrale  \\ 
DE09 & Aareal Bank AG  \\ 
DE10 & Deutsche Apotheker- und \"Arztebank eG  \\ 
DE11 & HASPA Finanzholding  \\ 
DE14 & Landeskreditbank Baden-W\"urttemberg-F\"orderbank  \\ 
DE15 & Landwirtschaftliche Rentenbank  \\ 
DE16 & M\"unchener Hypothekenbank eG  \\ 
DE20 & DZ Bank AG Deutsche Zentral-Genossenschaftsbank  \\ 
DE25 & Deutsche Pfandbriefbank AG  \\ 
DE26 & Erwerbsgesellschaft der S-Finanzgruppe mbH \& Co. KG  \\ 
DE27 & HSH Beteiligungs Management GmbH  \\ 
DE28 & State Street Europe Holdings Germany S.\`a.r.l. \& Co. KG  \\ 
DE29 & Volkswagen Bank GmbH \\ 
DK01 & Danske Bank  \\ 
DK02 & Jyske Bank  \\ 
DK03 & Sydbank  \\ 
DK05 & Nykredit Realkredit  \\ 
EE01 & AS LHV Group  \\ 
ES01 & Banco Santander * \\ 
ES02 & Banco Bilbao Vizcaya Argentaria  \\ 
ES03 & Banco de Sabadell  \\ 
ES04 & Banco Financiero y de Ahorros  \\ 
ES07 & Caja de Ahorros y M.P. de Zaragoza  \\ 
ES08 & Kutxabank  \\ 
ES09 & Liberbank  \\ 
ES11 & MPCA Ronda  \\ 
ES12 & Caja de Ahorros y Pensiones de Barcelona  \\ 
ES15 & Bankinter  \\ 
ES18 & Abanca Holding Financiero, S.A.  \\ 
ES19 & Banco de Cr\'edito Social Cooperativo, S.A.  \\ 
FI01 & OP-Pohjola Group  \\ 
FI02 & Kuntarahoitus Oyj  \\ 
FR01 & La Banque Postale  \\ 
FR02 & BNP Paribas * \\ 
FR03 & Soci\'et\'e G\'en\'erale * \\ 
FR06 & C.R.H. - Caisse de Refinancement de l'Habitat  \\ 
FR08 & RCI Banque  \\ 
FR09 & Soci\'et\'e de Financement Local  \\ 
FR12 & Groupe Cr\'edit Mutuel  \\ 
FR13 & Banque Centrale de Compensation (LCH Clearnet)  \\ 
FR14 & Bpifrance (Banque Publique d'Investissement)  \\ 
FR15 & Groupe BPCE * \\ 
FR16 & Groupe Cr\'edit Agricole * 
\end{tabular}
\begin{tabular}{ll}
GR01 & Eurobank Ergasias  \\ 
GR02 & National Bank of Greece  \\ 
GR03 & Alpha Bank  \\ 
GR04 & Piraeus Bank  \\ 
HU01 & OTP Bank Ltd \\
IE04 & AIB Group plc  \\ 
IE05 & Bank of Ireland Group plc  \\ 
IE06 & Citibank Holdings Ireland Limited  \\ 
IE07 & DEPFA BANK Plc  \\ 
IS01 & Arion banki hf  \\ 
IS02 & \'Islandsbanki hf.  \\ 
IS03 & Landsbankinn  \\ 
IT01 & Intesa Sanpaolo S.p.A.  \\ 
IT02 & UniCredit S.p.A. * \\ 
IT03 & Banca Monte dei Paschi di Siena S.p.A.  \\ 
IT04 & Unione Di Banche Italiane Societ\`a Cooperativa Per Azioni  \\ 
IT05 & Banca Carige S.P.A. - Cassa di Risparmio di Genova e Imperia  \\ 
IT07 & Banca Popolare Dell'Emilia Romagna - Societ\`a Cooperativa  \\ 
IT09 & Banca Popolare di Sondrio  \\ 
IT13 & Mediobanca - Banca di Credito Finanziario S.p.A.  \\ 
IT16 & Banco BPM Gruppo Bancario  \\ 
IT17 & Credito Emiliano Holding SpA  \\ 
IT18 & Iccrea Banca Spa Istituto Centrale del Credito Cooperativo  \\ 
LU01 & Banque et Caisse d'Epargne de l'Etat  \\ 
LU02 & Precision Capital S.A.  \\ 
LU03 & J.P. Morgan Bank Luxembourg S.A.  \\ 
LU04 & RBC Investor Services Bank S.A.  \\ 
LU05 & State Street Bank Luxembourg S.A.  \\ 
MT01 & Bank of Valletta plc  \\ 
MT02 & Commbank Europe Ltd  \\ 
MT03 & MDB Group Limited  \\ 
NL01 & Bank Nederlandse Gemeenten N.V.  \\ 
NL02 & Co\"operatieve Centrale Raiffeisen-Boerenleenbank B.A.  \\ 
NL03 & Nederlandse Waterschapsbank N.V.  \\ 
NL07 & ABN AMRO Group N.V.  \\ 
NL08 & ING Groep N.V. * \\ 
NL09 & Volksholding B.V.  \\ 
NO01 & DNB Bank Group  \\ 
NO02 & SPAREBANK 1 SMN   \\ 
NO03 & SR-bank  \\ 
PL01 & PKO BANK POLSKI  \\ 
PL07 & Bank Polska Kasa Opieki SA  \\ 
PT01 & Caixa Geral de Dep\'ositos  \\ 
PT02 & Banco Comercial Portugu\^es  \\ 
PT04 & Caixa Central de Cr\'edito Agr\'icola M\'utuo, CRL  \\ 
PT05 & Caixa Econ\'omica Montepio Geral, Caixa Econ\'omica Banc\'aria SA  \\ 
PT06 & Novo Banco, SA  \\ 
RO01 & Banca Transilvania  \\ 
SE01 & Nordea Bank AB (publ) $\dagger$\\ 
SE02 & Skandinaviska Enskilda Banken AB (publ) (SEB)  \\ 
SE03 & Svenska Handelsbanken AB (publ)  \\ 
SE04 & Swedbank AB (publ)  \\ 
SE05 & Kommuninvest - group  \\ 
SE06 & L\"ansf\"ors\"akringar Bank AB - group  \\ 
SE07 & SBAB Bank AB - group  \\ 
SI02 & Nova Ljubljanska banka d. d.  \\ 
SI04 & Abanka d.d.  \\ 
SI05 & Biser Topco S.\`a.r.l.  \\ 
UK01 & Royal Bank of Scotland Group plc $\dagger$\\ 
UK02 & HSBC Holdings plc * \\ 
UK03 & Barclays plc * \\ 
UK04 & Lloyds Banking Group plc  \\ 
UK05 & Nationwide Building Society  \\ 
UK06 & Standard Chartered Plc *
\end{tabular}
\caption{\label{tab:banks_full} IDs and full names of banks in the 2018 EBA Network. Banks marked by asterisk (*) were G-SIBs in all years (2014, 2016, 2018); banks marked by dagger ($\dagger$) were G-SIBs in 2014 and 2016, but not in 2018.}
\end{table}

\end{appendix}

\end{document}